\documentclass{ws-procs10x7}

\begin{document}

\title{Review of Heavy Quark Physics -- Theory}

\author{Ahmed Ali}

\address{Deutsches Elektronen-Synchrotron DESY, Hamburg,
Germany\\E-mail: ahmed.ali@desy.de}

\twocolumn[\maketitle\abstract{
Recent progress in the theory of $B$-meson decays is reviewed with emphasis on
the aspects related to the $B$-factory data.
}]

\section{Introduction}

The two $B$-meson factories operating at the
KEK and SLAC $e^+e^-$ storage rings have outperformed their projected luminosities and have
produced a wealth of data, in particular on CP asymmetries and rare
 $B$-decays~\cite{Sakai-ichep04,Giorgi-ichep04}.
Impressive results were also presented at this conference by the two
Fermilab experiments CDF and D0 in B physics~\cite{Herndon-ichep04} and top-quark
physics~\cite{Denisov-ichep04}. These and other experiments~\cite{Patera-ichep04} have provided
precision measurements of the Cabibbo-Kobayashi-Maskawa (CKM)~\cite{Cabibbo:yz,Kobayashi:fv}
matrix elements establishing the unitarity of this matrix~\cite{Ali:2003te}.
The KM-mechanism~\cite{Kobayashi:fv} of CP violation 
is now being tested with ever increasing precision in a large number of
 decay modes~\cite{pdg:2004,Ligeti-ichep04}.

 In my talk I  will concentrate on the following three topics:
\vspace*{-0.1cm}
\begin{itemize}
\item     
Current determination of $\vert V_{cb}\vert$ and  $\vert V_{ub}\vert$.
\item Progress in flavor-changing-neutral current (FCNC) induced rare 
 $B$-decays.
\item Comparison of various theoretical approaches to non-leptonic $B$-decays with
       current data in selected two-body $B$-decays.
 \end{itemize}
\vspace*{-0.1cm}
 Aspects of  charm physics are discussed
in the plenary talks by
Ian Shipsey~\cite{Shipsey-ichep04}, and  the Lattice-QCD results related to 
flavor physics are taken up by Shoji Hashimoto~\cite{Hashimoto-ichep04}. 
\section{Current determinations of  $\vert V_{cb}\vert$ and  $\vert V_{ub}\vert$}

The CKM matrix is written below in the
Wolfenstein parameterization~\cite{Wolfenstein:1983yz} in terms of the 
four parameters: $A, ~\lambda,~~\rho, ~~\eta$:
\begin{small}
\begin{eqnarray}
V_{\rm CKM} &\equiv& \nonumber\\
& &\hspace*{-1.5cm}
 \left(\matrix{
 1-{1\over 2}\lambda^2 & \lambda
 & \hspace*{-0.7cm}A\lambda^3 \left( \rho - i\eta \right) \cr
 -\lambda ( 1 + i A^2 \lambda^4 \eta )
& 1-{1\over 2}\lambda^2 &  \hspace*{-0.7cm}A\lambda^2 \cr
 A\lambda^3\left(1 - \rho - i \eta\right) & -A\lambda^2\left(1+i \lambda^2 \eta\right) 
& \hspace*{-0.7cm}1 
\cr}\right).
\nonumber
\label{CKM-W}
\end{eqnarray}
\end{small}
 Anticipating precision data, a perturbatively improved version~\cite{Buras:1994ec} of
the Wolfenstein parameterization will be used below with
 $\bar \rho =\rho(1-\lambda^2/2),~~\bar \eta= \eta(1-\lambda^2/2)$.

 Unitarity of the CKM matrix
implies six  relations, of which the one resulting from the
equation $V_{ud}V_{ub}^* + V_{cd}V_{cb}^* + V_{td} V_{tb}^*=0$ is the principal focus of the
current experiments in $B$-decays.
This is a triangle relation in the complex plane 
(i.e.\ $\bar{\rho}$--$\bar{\eta}$ space), and  the three angles of this 
triangle are called $\alpha$, $\beta$ and $\gamma$, with 
the BELLE convention being $\phi_2=\alpha$, $\phi_1=\beta$ and 
$\phi_3=\gamma$. The unitarity relation in discussion can also be written as
\begin{equation}
R_b {\rm e}^{i\gamma} + R_t {\rm e}^{-i \beta} =1\,,
\label{eq:trianglerel} 
\end{equation}
where
$R_b = \left (1-
\frac{\lambda^2}{2}\right )\, \frac{1}{\lambda} \,\left \vert 
\frac{V_{ub}}{V_{cb}} \right \vert 
=\sqrt{\bar{\rho}^2 + \bar{\eta}^2}$ and
$R_t =\frac{1}{\lambda} \, \left \vert 
\frac{V_{td}}{V_{cb}} \right \vert =\sqrt{(1-\bar{\rho})^2 + \bar{\eta}^2}$.
Thus, precise determination of $\vert V_{cb}\vert$, $\vert V_{ub} \vert$ and $\vert V_{td} \vert$
and the three CP-violating phases $\alpha$, $\beta$, $\gamma$  is
crucial in testing the CKM paradigm.

\subsection{$\vert V_{cb} \vert$ from the decays $B~\to~X_c~\ell~\nu_\ell$}
\label{subsec:vcbincl}

 Determinations of $\vert V_{cb}\vert $ are
based on the semi-leptonic decay  $b \to c \ell \nu_\ell$. This
transition can be measured either inclusively through the process $B \to X_c \ell \nu_\ell$, where
$X_c$ is a hadronic state with a net $c$-quantum number, or exclusively, such as the decays
$B \to (D,D^*) \ell \nu_\ell$. In either case, intimate knowledge of QCD is required to go from
the partonic process to the hadronic states. The fact that $m_b \gg \Lambda_{\rm QCD}$ has led to
novel applications of QCD in which heavy quark expansion (HQE) plays a central
 role~\cite{Manohar:dt} and the underlying theory
is termed as HQET. Concentrating on the inclusive decays,
the semi-leptonic decay rate can be calculated as a power series
\begin{equation}
\Gamma = \Gamma_0 + \frac{1}{m_b} \Gamma_1 + \frac{1}{m_b^2} \Gamma_2 + \frac{1}{m_b^3} \Gamma_3
+...,  
\end{equation}
where each $\Gamma_i$ is a perturbation series in $\alpha_s(m_b)$, the QCD coupling constant
at the scale $m_b$. Here $\Gamma_0 $
is the decay width of a free $b$-quark, which gives the parton-model result. The coefficient of
the leading power correction $\Gamma_1 $ is absent~\cite{CGG}, and the
effect of the $1/m_b^2$ correction is collected in $\Gamma_2$, which can be expressed in terms of two
non-perturbative parameters called $\lambda_1$ - kinetic energy of the $b$-quark - and
$\lambda_2$ - its chromomagnetic moment. These quantities, also called
$\mu_\pi^2$ and $\mu_G^2$, respectively, in the literature, are defined in terms of the following
matrix elements~\cite{incl,MaWi,Blok:1993va,Mannel:1993su}:
\begin{eqnarray}
2 M_B \lambda_1 &\equiv& \langle B (v) 
                | \bar{Q}_v  (iD)^2  Q_v|  B (v) \rangle ,\\
6 M_B \lambda_2  &\equiv&
\langle B (v)   | \bar{Q}_v \sigma_{\mu \nu} [iD^\mu , iD^\nu]
Q_v | B (v) \rangle~,
 \nonumber
\end{eqnarray}
where $D_\mu$ is the covariant derivative and heavy quark fields are characterized by
the 4-velocity, $v$.
At ${\cal O} (\Lambda_{\rm QCD}^3/m_b^3)$,  six new matrix elements enter in $\Gamma_3 $,
usually denoted by $\rho_{1,2}$ and ${\cal T}_{1,2,3,4}$, discussed below.

Data have been analyzed in the theoretical accuracy in which corrections up to 
${\cal O}(\alpha_s^2 \beta_0)$, ${\cal O}(\alpha_s \Lambda_{\rm QCD}/m_b)$ and 
${\cal O}(\Lambda_{\rm QCD}^3/m_b^3)$  are taken into
 account~\cite{Bauer:2004ve,Gambino:2004qm}, with $\beta_0$ being the lowest oder $\beta$-function in QCD.
 The choice of the parameters entering the fit depends on whether or
not an expansion in $1/m_c$ is performed. 
 In addition to this choice, a quark mass scheme has to be specified.
 Bauer et al.~\cite{Bauer:2004ve} have carried out a comprehensive
 study of both of these issues using
five quark mass schemes: $1S$, PS, $\overline{\rm MS}$, kinematic, and the pole mass.

To extract the value of $\vert V_{cb}\vert$ and other fit parameters,  three different
distributions, namely the charged lepton energy spectrum and the hadronic invariant mass spectrum in
$B \to X_c \ell \bar {\nu}_\ell$, and the photon energy spectrum in $B \to X_s \gamma$ have
been studied.
Theoretical analyses are carried out in terms of the moments and not the distributions themselves.
 Defining the integral
$R_n(E_{\rm cut}, \mu) \equiv \int_{E_{\rm cut}} dE_\ell (E_\ell -\mu)^n d\Gamma/dE_\ell$,
where $E_{\rm cut}$ is a lower cut on the charged lepton energy, moments of
the lepton energy spectrum are given by $\langle E_\ell^n \rangle = R_n(E_{\rm cut}, 0)/R_0(E_{\rm cut},0)$.
For the $B \to X_c \ell \bar{\nu}_\ell$ hadronic invariant mass spectrum, 
the moments are defined likewise with the cutoff $E_{\rm cut}$.
Analyses of the $B$-factory data have been presented at this conference by the
BABAR~\cite{Luth-ichep04,Aubert:2004td,Aubert:2004te} and BELLE~\cite{Abe:2004zv,Abe:2004ks}
 collaborations. Studies along these lines of some of the moments
were also undertaken by the CDF~\cite{CDF-had-mom-2004}, CLEO~\cite{Mahmood:2002tt,Mahmood:2004kq} and
DELPHI~\cite{DELPHI-03} collaborations.
 
The  BABAR collaboration have studied the dependence of the lepton and hadron moments
on the cutoff $E_{\rm cut}$ and compared their measurements with the theoretical calculation
by Gambino and Uraltsev~\cite{Gambino:2004qm} using the so-called kinematic scheme for the $b$-quark mass
 $m_b^{\rm kin} (\mu)$, renormalized at the scale $\mu=1$ GeV.
%
%
 Excellent agreement between experiment and theory
is observed, allowing to determine the fit parameters in this scheme with the
 results~\cite{Luth-ichep04,Aubert:2004te}:  
\begin{eqnarray}
\nonumber
\vspace*{-24mm}
\hspace*{-15mm} & &  |V_{cb}| = \nonumber\\
& &(41.4 \pm 0.4_{exp} \pm 0.4_{HQE} \pm 0.6_{th})\, \times 10^{-3},  
\nonumber \\
& &  m_b~(1 {\rm GeV}) = 
\nonumber  \\
& & \hspace*{1mm} 
(4.61 \pm 0.05_{exp} \pm 0.04_{HQE} \pm 0.02_{th})~{\rm GeV}, 
\nonumber \\
& &  m_c~(1 {\rm GeV}) = 
\nonumber  \\
& & \hspace*{1mm} 
(1.18 \pm 0.07_{exp} \pm 0.06_{HQE} 
\pm 0.02_{th})~{\rm GeV}. \nonumber\\ 
\label{eq:vcb-babar}
\end{eqnarray}
\vspace*{-1mm}
The  global fit of the data
 from the BABAR, BELLE, CDF, CLEO and DELPHI
collaborations in the so-called $1S$-scheme for the $b$-quark mass 
undertaken  by Bauer et al.~\cite{Bauer:2004ve}
leads to the following fit values for $\vert V_{cb} \vert$ and $m_b$:
%
%
%
%
%
\begin{eqnarray}
& &  |V_{cb}| = \left( 41.4 \pm 0.6 \pm 0.1_{\tau_B}\right) \times 10^{-3}, 
\nonumber \\[4mm]
& &  m_b^{1S} = (4.68 \pm 0.03)\,\mbox{GeV}.
\label{eq:vcb-bauer}
\end{eqnarray}
The two analyses yielding (\ref{eq:vcb-babar}) and (\ref{eq:vcb-bauer}) are in excellent agreement
 with each other. The achieved accuracy
$\delta \vert V_{cb}\vert/\vert V_{cb} \vert \simeq 2\%$ is impressive, and
 the precision on $m_b$ is also remarkable,
$\delta m_b/m_b =O(10^{-3})$, with a similar precision obtained on the mass difference $m_b-m_c$.
\subsection{$\vert V_{cb} \vert$ from $B \to(D,D^*) \ell \nu_\ell$ decays}
\label{subsec:vcbexcl}
The classic application of HQET in heavy $\to$ heavy decays is 
$B \to D^* \ell \nu_\ell$. The differential distribution in the variable $\omega (=v_B.v_{D^*})$,
where $v_B (v_{D^*})$ is the four-velocity of the $B(D^*)$-meson, is given by
\begin{eqnarray}
\frac{d\Gamma}{d\omega} &=& \frac{G_F^2}{4 \pi^3}  \,  \vert V_{cb} \vert^2 \, m_{D^*}^3 \, 
(m_B - m_{D^*})^2 \, \nonumber\\
& & \times \left (\omega^2 -1 \right )^{1/2} \,{\cal G}(\omega) \, \vert {\cal F} (\omega) \vert^2~,
\nonumber
\end{eqnarray}
where ${\cal G}(\omega)$ is a phase space factor with 
 ${\cal G}(1)=1$, and 
 ${\cal F}(\omega)$ is the Isgur--Wise (IW) function~\cite{Isgur:vq}
 with the normalization at the symmetry point
${\cal F}(1)=1$.  Leading $\Lambda_{\rm QCD}/m_b$ corrections in ${\cal F}(1)$  
 are absent  due to Luke's theorem~\cite{Luke:1990eg}. Theoretical issues are the precise determination of the 
second order power correction to ${\cal F}(\omega=1)$, the
slope  $\rho^2$ and the curvature $c$ of the IW-function:
\begin{equation}
{\cal F}(\omega) ={\cal F}(1)\, \left [1 + \rho^2\,(\omega -1) + 
c\,(\omega -1)^2 + ...\right ].
\nonumber
\end{equation}
Bounds on $\rho^2$ have been obtained by Bjorken~\cite{Bjorken:1990hs} and Uraltsev~\cite{Uraltsev:2000ce},
which can be combined to yield $\rho^2 > 3/4$. Likewise,
 bounds on the second (and higher) derivatives of the IW-function have been worked out
 by the Orsay group~\cite{Jugeau:2004qd}, yielding $c > 15/32$~\cite{Oliver-ichep04}.
These bounds have not been used (at least not uniformly) in the current analyses of the
$B \to D^* \ell \nu_\ell$ data by the experimental groups. This, combined with the possibility
that the data sets  may also differ significantly from experiment to experiment, results in considerable dispersion in
the values of  ${\cal F}(1)\vert V_{cb}\vert$ and $\rho^2$ and hence in a large $\chi^2$ of the combined fit,
summarized by the HFAG averages~\cite{hfag04}:
\begin{eqnarray}
{\cal F}(1) \vert V_{cb} \vert &=&(37.7 \pm 0.9) \times 10^{-3}~,\\
\rho^2&=&1.56 \pm 0.14 \quad (\chi^2=26.9/14).
\nonumber
\end{eqnarray}
To convert this into a value of $\vert V_{cb} \vert$, we need to know 
${\cal F}(1)$. 
In terms of the perturbative (QED and QCD) and non-perturbative (leading $\delta_{1/m^2}$ and sub-leading
$\delta_{1/m^3}$) corrections, ${\cal F}(1)$ can be expressed as follows:
\begin{equation}
{\cal F}(1)= \eta_A\left[ 1 + \delta_{1/m^2} + \delta_{1/m^3} \right]~,
\end{equation}
where $\eta_A$ is the perturbative renormalization of the IW-function, known in the meanwhile to three
loops~\cite{Archambault:2004zs}. One- and two-loop corrections yield $\eta_A\simeq 0.933$
and the $O(\alpha_s^3)$
contribution amounts to $\eta_A^{(3)}=-0.005$.
 Default value  of ${\cal F}(1) $ used by HFAG is based on the BABAR~book~\cite{BABARbook}
 ${\cal F}(1)= 0.91 \pm 0.04$. A recent Lattice-QCD calculation in the quenched approximation yields~\cite{Hashimoto:2001nb}
 $ {\cal F}(1)= 0.913^{+0.0238~+0.0171}_{-0.0173~-0.0302}$, which is now being reevaluated with
dynamical quarks~\cite{Hashimoto-ichep04}.
 With  ${\cal F}(1)=0.91\pm 0.04$, HFAG quotes the following average~\cite{hfag04} 
\begin{equation}
\vert V_{cb} \vert_{B \to D^* \ell \nu_\ell} = 
(41.4 \pm 1.0_{\rm exp} \pm 1.8_{\rm  theo})\times 10^{-3}.
\label{eq:vcb-excl}
\end{equation}
The resulting value of $\vert V_{cb} \vert $ is in excellent agreement with the
ones given in (\ref{eq:vcb-babar}) and (\ref{eq:vcb-bauer}) obtained 
from the inclusive decays. However, in view of the rather large $\chi^2$ of the
fit and the remark on the slope of the IW-function made earlier, there is room for
systematic improvements in the determination of $\vert V_{cb}\vert$ from the
exclusive analysis.  

The decay $B \to D \ell \nu_\ell$ still suffers from paucity of data. An analysis of the current data
in this decay mode using the HQET formalism, which admits leading $1/m_b$ corrections, is~\cite{hfag04}
$\vert V_{cb} \vert = (40.4 \pm 3.6_{\rm exp} \pm 2.3_{\rm th})\times 10^{-3}$. 
 The determination of $\vert V_{cb} \vert $ from $B \to D \ell \nu_\ell$
can be significantly improved at the B-meson factories. 

\vspace*{-0.6cm}
\subsection{$\vert V_{ub} \vert$ from the decays $B \to X_u \ell \nu_\ell$}
\label{subsec:vubincl}

HQET techniques allow to calculate the inclusive decay rate $B \to X_u \ell \nu_\ell$ rather
accurately. However, the experimental problem in measuring this transition lies in the huge background from the dominant decays
$B \to X_c \ell \nu_\ell$ which can be brought under control only through severe cuts on the
kinematics. For example, these cuts are imposed on the lepton energy,
demanding $E_\ell > (m_B^2-m_D^2)/2m_B$, and/or 
the momentum transfer to the lepton pair $q^2$ restricting it below a threshold value $q^2 < q^2_{\rm max}$,
and/or the hadron mass recoiling against the leptons, which is required to satisfy
$m_X < m_D$. With these cuts,
the phase space of the decay $B \to X_u \ell \nu_\ell$ is  greatly reduced. A bigger problem is encountered
in the end-point region (also called the shape function region), where the leading power correction is no
longer $1/m_b^2$ but rather $1/m_b\Lambda_{\rm QCD}$, slowing the convergence of the expansion. 
Moreover, in the region of energetic leptons with low invariant mass hadronic states,
 $E_X \sim m_b, ~m_X^2 \sim m_D^2 \sim \Lambda_{\rm QCD} m_b \ll m_b^2$, the differential rate is sensitive
to the details of the shape function $f(k_+)$~\cite{Luke:2003nu}, where $ k_{+}=k_0+k_3$
with $k^\mu \sim O(\Lambda_{\rm QCD})$.

The need to know $f(k_+)$ can be circumvented by doing a combined analysis of the data
on $B \to X_u \ell \nu_\ell$ and $B \to X_s \gamma$. Using 
the operator product expansion (OPE) to calculate the photon energy
spectrum in the inclusive decay $B \to X_s \gamma$,  
the leading terms in the spectrum (neglecting the bremsstrahlung corrections) can be re-summed into a
 shape function~\cite{shape}:
\begin{equation}
\frac{d\Gamma_s}{dx} = \frac{G_F^2 \alpha m_b^5}{32 \pi^4}\, |
          V_{ts} V_{tb}^*|^2\, |C_7^{\rm eff}|^2\, f(1-x) ~,
\end{equation}
where $x = \frac{2E_\gamma}{m_b}$. In the leading order, $E_\ell$- and $M_{X_u}$ spectra in
$B \to X_u \ell \nu_\ell$ are also governed by $f(x)$. Thus, 
$f(x)$ can be measured in $B \to X_s \gamma$ and used
in the analysis of data in $B \to X_u \ell \nu_\ell$.
  
 Following this argument,
 a useful relation emerges~\cite{Leibovich:1999xf,Neubert:2001sk,Bauer:2001mh}
\begin{equation}
\vert \frac{V_{ub}}{V_{tb}V_{ts}^*}\vert =\left( \frac{3 \alpha}{\pi} \vert C_7^{\rm eff}\vert^2
\frac{\Gamma_u(E_c)}{\Gamma_s(E_c)}\right)^{\frac{1}{2}}( 1 + \delta(E_c))~,
\label{eq:vubsgincl}
\end{equation}  
where 
\begin{eqnarray}
\Gamma_u(E_c) \equiv \int_{E_c}^{m_B/2} d E_\ell \frac{d\Gamma_u}{dE_\ell}~, \nonumber\\
& &\hspace*{-4.5cm}\Gamma_s(E_c) \equiv \frac{2}{m_b} \int_{E_c}^{m_B/2}
 d E_\gamma (E_\gamma -E_c)
 \frac{d\Gamma_s}{dE_\gamma}~,\nonumber\\
\end{eqnarray}
and $\delta(E_c)$ incorporates the sub-leading terms in ${\cal O}(\Lambda_{\rm QCD}/m_b)$,
which can only be modeled at present.
 In addition, there are perturbative corrections to the
spectra and in the relation (\ref{eq:vubsgincl})~\cite{shape,Korchemsky:1994jb,Leibovich:1999xf}.

Theoretical uncertainties in the extraction of $\vert V_{ub} \vert$ arise from the
weak annihilation (WA) contributions~\cite{Bigi:1993bh,Voloshin:2001xi} which
depend on the size of factorization violation. Also, the $O(\Lambda_{\rm QCD}^3/m_b^3)$
contributions,  which have  been studied using a  model~\cite{DeFazio:1999sv}
for $f(x)$, are found to
grow as $q^2$ and $m_{\rm cut}$ are increased.

At this conference, the BELLE collaboration~\cite{Abe:2004zm} have presented impressive new analyses 
for the inclusive $B \to X_u \ell \nu_\ell$ decays with fully reconstructed tags.
This has allowed them to measure the partial branching ratio (with $m_X < 1.7$ GeV, $q^2 > 8$ GeV$^2$)
\begin{eqnarray}
\Delta {\cal B}(B \to X_u \ell \nu_\ell) = \left( 0.99 \pm 0.15 ({\rm stat})
\pm 0.18\right.\nonumber\\
& & \hspace*{-7.0cm} \left. ({\rm syst.}) \pm 0.04 (b \to u) \pm 0.07 (b \to c)\right)\times 10^{-3}~.
\nonumber\\
\vspace*{-0.8cm}
\end{eqnarray}
To get the full branching ratio, a knowledge of the shape function is needed which is obtained in a 
model-dependent analysis~\cite{Limosani:2004jk} from the measured $B \to X_s \gamma$ spectrum.
 Using the expression
$ {\cal B}(B \to X_u \ell \nu_\ell) = \Delta B (B \to X_u \ell \nu_\ell)/f_u$, the BELLE analysis
 estimates $f_u=0.294\pm 0.044$~\cite{Abe:2004zm}.
Combined with the PDG prescription~\cite{pdg:2004} 
\begin{equation}
\vert V_{ub} \vert = 0.00424 \left[ \frac{B(B \to X_u \ell \nu_\ell)}{0.002}\frac{ 1.61~{\rm ps}}{\tau_B}\right],
\end{equation}
yields~\cite{Abe:2004zm} 
\begin{eqnarray}
\vert V_{ub} \vert &=& \left( 5.54 \pm 0.42 ({\rm stat}) \pm 0.50 ({\rm syst.})
\right.\nonumber\\
& & \hspace*{-0.6cm} \left. \pm 0.12 (b \to u) \pm 0.19 (b \to c)
\pm 0.42 (f_u)\right.\nonumber\\
 & & \hspace*{-0.6cm} \left. \pm 0.27 ( {\cal B} \to \vert V_{ub} \vert)\right)\times 10^{-3}.
\end{eqnarray} 
The corresponding determination of $\vert V_{ub} \vert$ by the BABAR collaboration using this method 
gives~\cite{Aubert:2004bq}
\begin{equation}
\vert V_{ub} \vert = (5.18 \pm 0.52 \pm 0.42) \times 10^{-3}~,
\end{equation}
where the errors are statistical and systematic, respectively. 
The current determination of $\vert V_{ub} \vert$ from inclusive
measurements including the above BABAR and BELLE measurements is summarized 
by  HFAG (Summer 2004 update)~\cite{hfag04}, with the average
\begin{equation}
\vert V_{ub} \vert=(4.70 \pm 0.44) \times 10^{-3}~,
\end{equation}
having a $\chi^2/$ p.d.f $= 6.7/7$. This amounts to about 10\% precision on $\vert V_{ub}\vert$.
Recently, a new method to determine $\vert V_{ub} \vert$ from the inclusive decays $B \to X_u \ell \nu_\ell$
has been proposed~\cite{Bosch:2004th} which uses a cut on the hadronic light-cone variable
 $P_+=E_X-\vert P_X\vert$. The efficiency and sensitivity to non-perturbative effects in the $P_+$-cut
method is argued to be similar to the one on the hadron mass cut, and the $P_+$-spectrum can be calculated in
a controlled theoretical framework. 

\subsection{$\vert V_{ub} \vert$ from exclusive decays}
 $\vert V_{ub} \vert$ has also been determined from the exclusive
decays $B \to (\pi,\rho) \ell \nu_\ell$. Theoretical accuracy is
limited by the imprecise knowledge of the form factors. 
A number of theoretical techniques has been used to determine them. These include,
among others,
Light-cone QCD sum rules~\cite{Colangelo:2000dp}, Quenched- and Unquenched-Lattice QCD
simulations~\cite{Hashimoto-ichep04}, and Lattice-QCD based 
phenomenological studies~\cite{Becirevic:2003}. New measurements and analysis 
of the decay $B \to \pi \ell \nu_\ell$ have been presented
at this conference by the BELLE collaboration and compared with a number of Lattice-QCD calculations,
and the extracted values of
$\vert V_{ub} \vert$ (in units of $10^{-3}$) are as follows~\cite{Ijima-ichep04}:\\
$
 |V_{ub}|_{\rm Quenched} = 
\left (3.90 \pm 0.71 \pm 0.23 ^{+0.62}_{-0.48}\right );\\ 
 |V_{ub}|_{\rm FNAL'04} =
\left (3.87 \pm 0.70 \pm 0.22 ^{+0.85}_{-0.51}\right ); \\
 |V_{ub}|_{\rm HPQCD} =
\left (4.73 \pm 0.85 \pm 0.27 ^{+0.74}_{-0.50}\right )$.
Hence, current Lattice-QCD results show considerable dispersion (about 20\%) in the extraction of
$\vert V_{ub} \vert$ from data.
%
%

To reduce the form-factor related uncertainties in extracting $\vert V_{ub} \vert$ from
exclusive decays $B \to (\pi,\rho) \ell \nu_\ell$, input from the rare $B$-decays
 $B \to (K,K^*) \ell^+\ell^-$ and HQET may be helpful.  A  proposal along these lines is
the so-called  Grinstein's double ratio which would
determine  $|V_{ub}|/|V_{tb} V_{ts}^*|$ from the end-point region of exclusive 
rare  $B$-meson decays~\cite{Grinstein-ichep04,Grinstein:2004vb}. To carry out this program one has to measure
four distributions: $B \to \rho \ell \nu_\ell$, $B \to K^* \ell^+\ell^-$, and $D \to (\rho,K^*) \ell \nu_\ell$.
With the help of this data and HQET, the ratio of the CKM factors  
$|V_{ub}|/|V_{tb} V_{ts}^*|$ can be determined through the double ratio 
\vspace*{-0.1cm}
\begin{equation}
\frac{\Gamma(\bar B \to \rho \ell \nu)}
     {\Gamma(\bar B \to K^* \ell^+ \ell^-)} \, 
\frac{\Gamma(D \to K^* \ell \nu)}
     {\Gamma(D \to \rho \ell \nu)}.
\vspace*{-0.1cm}
\end{equation}
At the $B$ factories, one expects enough data on these decays to allow a 10\% determination of $\vert V_{ub}\vert$
from  exclusive decays.
\vspace*{-0.3cm}
\section{Radiative, semileptonic and leptonic rare $B$ decays}

 Two inclusive rare $B$-decays of current experimental
interest are $B \to X_s \gamma$  and $B \to X_s l^+ l^-$,
where $X_s $ is any hadronic state with $s=1$, containing no charmed particles.
They probe the SM  in the electroweak $b \to s$ penguin sector.
 The CKM-suppressed decays $B \to X_d \gamma$
and $B \to X_d l^+ l^-$ are difficult to measure due to low rates and 
formidable backgrounds. Instead, the search for $b \to d$ radiative transitions has been
carried out in the modes $B \to (\rho,\omega) \gamma$ providing interesting constraints on
the CKM parameters. New and improved upper limits have been presented at this conference
on the branching ratio for $B_s^0 \to \mu^+\mu^-$, testing supersymmetry in the large-$\tan \beta$ domain.
We take up these decays below in turn.  

\subsection{$B \to X_s \gamma$: SM vs.~Experiments}
 The effective~Lagrangian~for the decays~$B \to X_s (\gamma, \ell^+\ell^-)$ in the SM
reads as follows:
\begin{eqnarray}
 \vspace*{-0.3cm}
  {\cal L_{\rm eff}} \;&=& \;\;  
\; \frac{4 G_F}{\sqrt{2}} V_{ts}^* V_{tb} \sum_{i=1}^{10} C_i(\mu) O_i~.
\label{eq:efflagrang}
\vspace*{-0.3cm}
\end{eqnarray}
 The operators and their Wilson coefficients
evaluated at the scale $\mu=m_b$ can be seen elsewhere~\cite{alimisiak:2003}.
QCD-improved calculations in the effective theory require three steps:\\
 (i) \underline{Matching} \, $C_i(\mu_0)$ ($\mu_0 \sim M_W, m_t$):
They have been calculated up to three loops~\cite{Bobek:199mk,Misiak:2004ew}.
 The three-loop matching is found to have  
less than $2\%$ effect on ${\cal B}(B \to X_s \gamma)$~\cite{Misiak:2004ew}.\\
 (ii) \underline{Operator mixing:} This involves calculation of the anomalous dimension matrix,
which is expanded in $\alpha_s(\mu)$.
 The anomalous dimensions up to  $\alpha_s^2(\mu)$ 
are known~\cite{Chetyrkin:1996vx} and  the $\alpha_s^3(\mu)$ calculations 
 are in progress.

\noindent
 (iii) \underline{Matrix elements} $\langle O_i \rangle (\mu_b)$ ($\mu_b \sim m_b$):
The first two terms in the expansion in
$\alpha_s(\mu_b)$ are known since long~\cite{Greub:1996tg}.
 The ${\cal O}(\alpha_s^2 n_f)$ part of the 3-loop calculations 
has recently been done by Bieri, Greub and Steinhauser~\cite{Bieri:2003ne}. The complete three-loop calculation
of $\langle O_i \rangle$, which is not yet in hand,  will reduce the quark mass scheme-dependence of the
branching ratio ${\cal B}(B \to X_s \gamma)$, and hence of the NLO decay rate for
 ${\cal B} (B \to X_s \gamma)$~\cite{Chetyrkin:1996vx,Gambino:2001ew}. Finally, one has to add the
Bremsstrahlung contribution $b \to s \gamma g$ to get the complete decay rate~\cite{Ali:1990tj}.

In the $\overline{\rm MS}$ scheme, the NLO branching ratio
 is calculated as~\cite{Gambino:2001ew,Buras:2002tp}:
\begin{equation}
{\cal B} (B \to X_s \gamma)_{\rm SM}= (3.70 \pm 0.30) \times 10^{-4}~.
\label{eq:misiak96}
\end{equation}
Including the uncertainty due to scheme-dependence, this amounts to a theoretical precision
of about $10\%$, comparable to the current experimental precision~\cite{hfag04} 
\begin{equation}
{\cal B} (B \to X_s \gamma)_{\rm Expt.}= (3.52 ^{+0.30}_{-0.28}) \times 10^{-4}~.
\label{eq:bsghfag04}
\end{equation}
Within stated errors, SM and data are in agreement.
In deriving (\ref{eq:misiak96}), unitarity of the CKM matrix yielding
 $\lambda_t=-\lambda_c=-A\lambda^2 +...=-(41.0 \pm 2.1)\times 10^{-3}$ has been used, where
 $\lambda_i=V_{ib} V_{is}^*$.
 The  measurement (\ref{eq:bsghfag04}) can also be used to determine
 $\lambda_t$.
Current data and the NLO calculations in the SM imply~\cite{alimisiak:2003}
\begin{equation}
\vert 1.69 \lambda_u + 
1.60 \lambda_c +0.60 \lambda_t \vert =
(0.94 \pm 0.07)\vert V_{cb} \vert~,
\end{equation}
leading to $\lambda_t= V_{tb}V^*_{ts} =-(47.0 \pm 8.0) \times 10^{-3}$. As $V_{tb}=1$
to a very high accuracy, ${\cal B} (B \to X_s \gamma)$ determines $V_{ts}$, both in sign
 and magnitude.

 The current (NLO) theoretical precision on ${\cal B} (B \to X_s \gamma)$ given in (\ref{eq:bsghfag04})
has recently been questioned~\cite{Neubert:2004dd}, using a multi-scale OPE
involving three low energy scales: $m_b$, $\sqrt{m_b \Delta}$ and $\Delta=m_b-2E_0$, where 
$E_0$ is the lower cut on the photon energy. With $E_0$ taken as $1.9$ GeV and $\Delta=1.1$ GeV,
one has considerable uncertainty in the decay rate due to the dependence on  $\Delta$.
%
%
\subsection{$B \to X_s  \ell^+ \ell^-$: SM vs.~Experiments}
The NNLO calculation of the decay   $B \to X_s l^+ l^-$ corresponds to the NLO
calculation of $B \to X_s \gamma$, as far as the
number of loops in the diagrams is concerned.  
%
%
%
%
 Including the leading power corrections in $1/m_b$ and $1/m_c$
and taking into account various parametric uncertainties, the branching ratios for the decays
$B \to X_s \ell^+ \ell^-$ in NNLO are~\cite{Ali:2002jg}:
\begin{eqnarray}
{\cal B}(B \to X_s e^+ e^-)_{\rm SM}  &\simeq& {\cal B}(B \to X_s \mu^+ \mu^-)_{\rm SM}
\nonumber\\
&&\hspace*{-1.3cm}= (4.2 \pm 0.7) \times 10^{-6}~,
\label{eq:aghl}
\end{eqnarray}
where a dilepton invariant mass cut, $m_{\ell \ell} > 0.2 $ GeV has been assumed for
comparison with data given below. These estimates make use of the NNLO calculation by
 Asatryan et al.~\cite{Asatryan:2001zw}, 
restricted to $\hat s \equiv q^2/m_b^2< 0.25$. The spectrum for 
 $\hat s> 0.25$ has been obtained from the NLO calculations using the scale $\mu_b \simeq m_b/2$,
as this choice of scale reduces the NNLO contributions. Subsequent NNLO calculations
cover the entire dilepton mass spectrum and are numerically in agreement with this
procedure, yielding~\cite{Ghinculov:2003qd,Bobeth:2003at}
${\cal B}(B \to X_s \mu^+ \mu^-)_{\rm SM}= (4.6 \pm 0.8) \times 10^{-6}$. The difference in the central
values in these results and (\ref{eq:aghl}) is of parametric origin.

 The BABAR and BELLE collaborations have measured the
invariant dilepton and hadron mass spectra in $B \to X_s \ell^+\ell^-$. Using the SM-based calculations to extrapolate
through the cut-regions, the current averages of the branching ratios are~\cite{hfag04}:
\begin{eqnarray}
{\cal B}(B \to X_s e^+ e^-)&=&~(4.70 ^{+1.24}_{-1.23}) \times 10^{-6},
\nonumber\\   
{\cal B}(B \to X_s \mu^+ \mu^-)&=& ~(4.26 ^{+1.18}_{-1.16})\times 10^{-6}~,
\nonumber\\ 
&&\hspace*{-3.2cm}{\cal B}(B \to X_s \ell^+ \ell^-)= ~(4.46 ^{+0.98}_{-0.96})\times 10^{-6}.
\label{eq:bsll-expt}
\end{eqnarray}
Thus, within the current experimental accuracy, which is typically 25\%, data and the SM
agree with each other in the $b \to s$ electroweak penguins. The measurements (\ref{eq:bsghfag04}) and
(\ref{eq:bsll-expt}) provide valuable constraints
 on beyond-the-SM physics scenarios.
 Following the earlier analysis to determine the Wilson coefficients  in $b \to s$
 transitions~\cite{Ali:1994bf,Ali:2002jg,Hiller:2003js},
it has been recently argued~\cite{Gambino:2004mv} that data now disfavor solutions in which the
 coefficient $C_7^{\rm eff}$ is similar in magnitude but opposite in sign to the SM coefficient.
For example, this constraint disfavors SUSY models with large $\tan \beta$ which admit
such solutions. 

Exclusive decays $B \to (K,K^*)\ell^+\ell^-$ $(\ell^\pm=e^\pm,\mu^\pm)$ have also been measured by the
BABAR and BELLE collaborations, and the current world averages of the branching ratios are~\cite{hfag04}:
\begin{eqnarray}
{\cal B}(B \to K \ell^+\ell^-)&=&~(5.74 ^{+0.71}_{-0.66}) \times 10^{-7},
\nonumber\\   
{\cal B}(B \to K^* e^+ e^-)&=& ~(14.4 ^{+3.5}_{-3.4})\times 10^{-7}~,
\nonumber\\ 
&&\hspace*{-3.2cm}{\cal B}(B \to K^* \mu^+ \mu^-)= ~(17.3 ^{+3.0}_{-2.7})\times 10^{-7}.
\label{eq:bkstll-expt}
\end{eqnarray}
They are also in agreement with the SM-based estimates of the same, posted as~\cite{Ali:2002jg}
${\cal B}(B \to K \ell^+ \ell^-)= (3.5 \pm 1.2)\times 10^{-7}$, 
${\cal B}(B \to K^* e^+ e^-)= (15.8 \pm 4.9)\times 10^{-7}$,
and  ${\cal B}(B \to K^* \mu^+ \mu^-)= (11.9 \pm 3.9)\times 10^{-7}$ with the error
dominated by uncertainties on the form factors~\cite{Ali:1999mm}.

 The Forward-backward (FB) asymmetry in the
decay $B \to X_s \ell^+ \ell^-$~\cite{Ali:1991is}, defined as 
\begin{eqnarray}
\bar {\cal A}_{\rm FB}(q^2) &=& \frac{1}{d {\cal B} ( B\to X_s \ell^+\ell^-) /d q^2  }
\\
 &&\hspace*{-1.9cm} \times \int_{-1}^1 d\cos\theta_\ell ~
 \frac{d^2 {\cal B} ( B\to X_s \ell^+\ell^-)}{d q^2  ~ d\cos\theta_\ell}
\mbox{sgn}(\cos\theta_\ell)~,\nonumber
\vspace*{-1.5cm} 
\end{eqnarray}
as well as the location of the zero-point of this asymmetry (called below
$q_0^2$) are precision tests of the SM. In NNLO, one has the following predictions:
 $q^2_0 =   (3.90 \pm 0.25)~{\rm GeV^2}~[(3.76 \pm 0.22_{\rm theory} \pm 0.24_{m_b})~{\rm GeV^2 }]$,
obtained by Ghinculov et al.~\cite{Ghinculov:2002pe} [Asatrian et al.~\cite{Asatrian:2002va}].
 In the SM (and its extensions in which the operator basis remains unchanged),
the  FB-asymmetry in $B \to K \ell^+ \ell^-$ is zero and in $B \to K^* \ell^+ \ell^-$
it  depends on the decay form factors. 
 Model-dependent studies yield small form factor-related
uncertainties in the zero-point of the asymmetry $\hat{s}_0=q_0^2/m_B^2$~\cite{Burdman:1998mk}.
 HQET provides a symmetry argument why the uncertainty in
$\hat{s}_0$ can be expected to be small which is determined by~\cite{Ali:1999mm} 
$ C_9^{eff}(\hat{s}_0) = -\frac{2 m_b}{M_B \hat{s}_0} C_7^{eff}$.
 However, 
$O(\alpha_s)$ corrections to the HQET-symmetry relations
lead to substantial change in the profile of the FB-asymmetry
function as well as a significant shift 
in $\hat{s}_0$~\cite{Beneke:2000wa,Beneke:2001at}. 
The zero of the FB-asymmetry is not very
precisely localized due to hadronic uncertainties, exemplified by the
estimate~\cite{Beneke:2000wa} $q_0^2=(4.2 \pm 0.6)$ GeV$^2$.  One also expects
that the intermediate scale $\Delta$-related uncertainties, worked out in the context of
 $B \to X_s \gamma$ and $B \to X_u \ell \nu_\ell$ in SCET~\cite{Neubert:2004dd,Lee:2004ja},
 will also renormalize the
dilepton spectra and the FB-asymmetries in $B \to (X_s,K^*,...) \ell^+\ell^-$. 

At this conference, BELLE have presented the
first measurement of the FB-asymmetry in the decays $B \to (K,K^*) \ell^+ \ell^-$.
 Data is compared with the SM predictions and with a
beyond-the-SM scenario in which the sign of $C_7^{\rm eff}$ is flipped, with no firm conclusions.
However, the beyond-the-SM scenario is disfavored on the grounds that it predicts too
high a branching ratio for $B \to X_s \ell^+\ell^- $ as well as for $B \to K^* \ell^+\ell^-$.
%
%
%

\vspace*{-0.3cm}
\subsection{$B \to V \gamma $: SM vs.~Experiments}
The decays $B \to V \gamma $ $(V= K^*,\rho,\omega $) have been calculated in the 
 NLO approximation using the effective Lagrangian given in (\ref{eq:efflagrang}) and its analogue for
 $b \to d$ transitions. Two dynamical approaches, namely the QCD
 Factorization~\cite{Beneke:1999br}
and  pQCD~\cite{Keum:2000ph}
 have been employed to establish factorization of the radiative decay amplitudes.
The QCD-F approach leads to the following factorization Ansatz for the
 $B \to V \gamma^{(*)}$ amplitude:
\begin{equation}
f_k (q^2) = C_{\perp k} \xi_\perp(q^2) +  C_{\| k} \xi_\|(q^2) +
\Phi_B \otimes T_k \otimes \Phi_V~,
\label{eq:qcdf-bkstar}
\end{equation}
where $f_k(q^2)$ is a form factor in the full QCD and the terms on the r.h.s. contain
factorizable and non-factorizable corrections. The functions $C_i$ $(i=\perp, \|)$ admit a
perturbative expansion  $C_{i} =C_{i}^{(0)} +\frac{\alpha_s}{\pi}C_{i}^{(1)}+...,$
with $C_{i}^{(0,1)}$ being the Wilson coefficients, and the so-called
 hard spectator corrections are given in the last term in (\ref{eq:qcdf-bkstar}).
 The symbol $\otimes$ denotes convolution of the perturbative QCD kernels with
the wave functions $\Phi_B$ and $\Phi_V$ of the $B$-Meson \& $V$-Meson. 
%
%
Concentrating first on the $B \to K^* \gamma$ decays, the branching ratio 
is enhanced in the NLO by a $K$-factor 
evaluated as~\cite{Beneke:2001at,Ali:2001ez,Bosch:2001gv}  
$1.5 \le K \le 1.7$.
The relation between $\xi_\perp^{(K^*)}(0) $ and the full QCD form factor
$ T_1^{K^*}(0) $ has been worked out in $O(\alpha_s)$ by Beneke and Feldmann~\cite{Beneke:2000wa}:
$ T_1^{K^*}(0) =(1 + O(\alpha_s))\xi_\perp^{(K^*)}(0)$.
Using the default values for  the $b$-quark mass in the pole mass scheme
$m_{b, {\rm pole}}=4.65~{\rm GeV}$
and the soft HQET form factor $\xi_\perp^{K^*}=0.35$ results in the following branching ratios~\cite{Ali:2004hn}
\begin{eqnarray}
{\cal B}_{\rm th} (B^0 \to K^{*0} \gamma) & \simeq & 
(6.9 \pm 1.1)\times 10^{-5} \,,
\nonumber \\[-1mm]
{\cal B}_{\rm th} (B^\pm \to K^{*\pm} \gamma) & \simeq &
 (7.4 \pm 1.2) \times 10^{-5}~.
\nonumber
\end{eqnarray}
The above theoretical branching ratios are to be compared with the current experimental
measurements~\cite{hfag04} 
\begin{eqnarray}
{\cal B}(B^0 \to K^{*0} \gamma) &=& (4.14 \pm 0.26) \times 10^{-5}; \nonumber\\ 
{\cal B}(B^\pm \to K^{*\pm} \gamma) &=&(3.98 \pm 0.35) \times 10^{-5}~.
\nonumber
\label{eq:bkstar-exp}
\end{eqnarray}
Consistency of the QCD-F approach with data requires
 $T_1^{K^*}(0)=0.27 \pm 0.02$. This is about 30\%
smaller than the typical estimates in QCD sum rules.

 In contrast to QCD-F, the pQCD approach is based on the so-called $k_\perp$-formalism,
in which the transverse momenta are treated in the Sudakov formalism. 
 Also, as opposed to the QCD-F approach,
in which the form factors are external input, pQCD calculates these form factors
yielding the following branching ratios~\cite{Keum:2004is}: 
 \begin{eqnarray}
{\cal B}(B^0\to K^{*0}\gamma)&=&(3.5^{+1.1}_{-0.8})\times 10^{-5}~,\nonumber\\
{\cal B}(B^{\pm}\to K^{*\pm}\gamma)&=&(3.4^{+1.2}_{-0.9})\times 10^{-5}~.
\nonumber
\end{eqnarray} 
The resulting form factor $T_1^{K^*}(0)=0.25 \pm 0.04$ is
in agreement with its estimate based on the QCD-F approach and data. 
%


The decays $B \to (\rho,\omega) \gamma$ involve in addition to the (short-distance) penguin
amplitude also significant long-distance
contributions, in particular in the decays $B^\pm \to \rho^\pm \gamma$.
 In the factorization approximation,  typical Annihilation-to-Penguin
amplitude ratio is estimated as~\cite{Ali:1995uy}:  
$\epsilon_{\rm A}(\rho^\pm \gamma)= 0.30 \pm 0.07$.
$O(\alpha_s)$ corrections to the annihilation amplitude in
$B^\pm \to \rho^\pm \gamma$ calculated in the 
leading-twist approximation  vanish in the chiral
limit~\cite{Grinstein:2000pc}. Hence,  non-factorizing annihilation
contributions are likely small which can be  tested experimentally in the decays 
$B^\pm \to \ell^\pm \nu_\ell \gamma$. The annihilation contribution to the decays
$B^0 \to \rho^0 \gamma$ and $B^0 \to \omega \gamma$  is expected to be suppressed 
(relative to the corresponding amplitdie in $B^\pm \to \rho^\pm \gamma$) due to   
the electric charges ($Q_d/Q_u=-1/2$) and the  colour factors, and the 
corresponding $A/P$ ratio for these decays is estimated as
$\epsilon_{\rm A}(\rho^0 \gamma) \simeq \epsilon_{\rm A}(\omega \gamma)
 \simeq 0.05$.

Theoretical branching ratios for $B \to (\rho,~\omega) \gamma$ decays can be more reliably
calculated in terms of the following ratios~\cite{Ali:2001ez,Ali:2004hn}:
\begin{equation}
R(\rho(\omega) \gamma) \equiv \frac{\overline {\cal B} (B \to \rho(\omega) \gamma)}
     {\overline {\cal B} (B \to K^* \gamma)}~.
\end{equation}
Including the $O(\alpha_s)$ and annihilation contributions~\cite{Ali:2004hn} 
$R (\rho^\pm/K^{*\pm}) = (3.3 \pm 1.0)\times 10^{-2}$ and 
$ R (\rho^0/K^{*0}) \simeq R (\omega/K^{*0})=(1.6\pm 0.5) \times 10^{-2}$. Using the well-measured
branching ratios $\overline {\cal B} (B \to K^* \gamma)$, and varying the CKM parameters in the allowed ranges,
one gets the following branching ratios~\cite{Ali:2004hn}:
$ \overline{\cal B}(B^\pm \to \rho^\pm \gamma) = (1.35 \pm 0.4)\times 10^{-6}$
and $ \overline{\cal B}(B^0 \to \rho^0\gamma) \simeq \overline{\cal B}(B^0 \to \omega
\gamma) = (0.65 \pm 0.2) \times 10^{-6}$.
To make comparison of the SM with the current
 data, the following averaged branching ratio is invoked
\begin{eqnarray}
\bar {\cal B} [B \to (\rho, \omega) \, \gamma] &\equiv& 
\frac{1}{2} \left \{ {\cal B} (B^+ \to \rho^+ \gamma) \right.\nonumber\\ 
+&& \hspace*{-2.8cm} \left. +\frac{\tau_{B^+}}{\tau_{B^0}} \left [ 
{\cal B} (B_d^0 \to \rho^0 \gamma) + 
{\cal B} (B_d^0 \to \omega \gamma) \right]   
\right \}.  
\nonumber
\end{eqnarray}
In terms of this averaged ratio, the current 
upper limits (at 90\% C.L.) are:
\begin{eqnarray}
\bar {\cal B}_{\rm exp} [B \to (\rho, \omega) \, \gamma] 
<1.4 \times 10^{-6}~~{\rm [BELLE]}~; \nonumber\\
&&\hspace*{-6.3cm} R[(\rho, \omega)/K^*] < 0.035;
~~|V_{td}/V_{ts}| < 0.22~,\nonumber\\ 
\end{eqnarray} 
and
\begin{eqnarray}
\bar {\cal B}_{\rm exp} [B \to (\rho, \omega) \, \gamma]
< 1.2 \times 10^{-6}~~{\rm [BABAR]};  \nonumber\\
&&\hspace*{-6.3cm} R[(\rho, \omega)/K^*] < 0.029;
~~|V_{td}/V_{ts}| < 0.19~.
  \nonumber\\
\end{eqnarray}
Constraints from the more stringent BABAR upper limit~\cite{Beryhill-ichep04}
$R[(\rho, \omega)/K^*] < 0.029$ on the
 CKM parameters  exclude up to almost 50\%
of the otherwise allowed parameter space, obtained  from the CKMfitter
group~\cite{Hocker:2001xe}.

%
%
\subsection{Current bounds on ${\cal B}(B^0_s \to \mu^+\mu^-$)}
New and improved upper limits have been presented at this conference by CDF and D0
 collaborations~\cite{Herndon-ichep04}
for the decays $B^0_s \to \mu^+\mu^-$ and $B^0_d \to \mu^+\mu^-$:
\begin{eqnarray}
{\cal B}(B^0_s \to \mu^+\mu^-) &<& 3.8~[5.8] \times 10^{-7} ~~{\rm D0 [CDF]}~,\nonumber\\
&&\hspace*{-3.0cm}{\cal B}(B^0_d \to \mu^+\mu^-) < 1.5 \times 10^{-7}~~{\rm [CDF]}.
\end{eqnarray}
The CDF and DO upper limits have been combined to yield 
${\cal B}(B_s^0 \to \mu^+\mu^-) < 2.7 \times 10^{-7}$, to be compared with the
SM predictions~\cite{Buchalla:1993bv}
 ${\cal B}(B^0_{s(d)} \to \mu^+\mu^-) = 3.4 \times 10^{-9} (1.0 \times 10^{-10})$ within
$\pm 15\%$ theoretical uncertainty.
Hence, currently there is no sensitivity for the SM decay rate.
 However, as the leptonic branching ratios
probe the Higgs sector in beyond-the-SM scenarios, such as supersymmetry, and depend sensitively on 
$\tan \beta$, the Tevatron upper limit on ${\cal B}(B^0_{s(d)} \to \mu^+\mu^-)$
 probes the large $\tan \beta$ (say, $> 50$)
parameter space, though the precise constraints are model dependent~\cite{Dermisek:2003vn,Ellis:2004tc}.

\section{$B \to M_1 M_2$ Decays}
Exclusive non-leptonic decays are the hardest nuts to crack in the theory of $B$-decays!
 Basically, there are four different theoretical approaches to
calculate and/or parameterize the hadronic matrix elements in $B \to M_1 M_2$ decays:
\begin{enumerate}
\item SU(2)/SU(3) symmetries and phenomenological Ansaetze
~\cite{Gronau:1990ka,Buras:1994pb,Deshpande:1994pw,Gronau:2002gj}

\item Dynamical approaches based on perturbative QCD, such as the QCD Factorization~\cite{Beneke:1999br}
and the competing pQCD approach~\cite{Keum:2000ph,Hnli-ichep04}. These techniques are very popular and have
a large following in China as well~\cite{Chang:2004cc}.

\item Charming Penguins~\cite{Ciuchini:2001gv,Pierini-ichep04} using the renomalization group
      invariant topological approach of Buras and Silvestrini~\cite{Buras:1998ra}.
 
\item Soft Collinear Effective Theory (SCET), for which several formulations
 exist. At this conference, SCET and its
applications are reviewed by Bauer, Pirjol, and
Stewart~\cite{Stewart-ichep04} to which we refer for detailed discussions.
\end{enumerate}

These approaches will be discussed on the example of the  decays $B \to \pi \pi$ and $B \to K\pi$ for which now
there exist enough data to extract the underlying dynamical parameters.

\subsection{$B \to \pi \pi$: SM vs.~Experiments}
  There are three dominant topologies
in the decays $B \to \pi \pi$ termed as Tree (T), Penguin (P) and Color-suppressed (C). In addition, there are
several other subdominant topologies which will be neglected in the discussion below. Parameterization of
the T, P, and C amplitudes is convention-dependent. In the Gronau-Rosner c-convention~\cite{Gronau:2002gj}, these
 amplitudes can be
represented as
\begin{eqnarray}
\sqrt 2 \, A^{+0} & = &  
- |T| \, {\rm e}^{i \delta_T} \, {\rm e}^{i \gamma} 
\left [ 1 + |C/T| \, {\rm e}^{i \Delta} \right ]~, 
\nonumber \\ 
A^{+-} & = &  
- |T| \, {\rm e}^{i \delta_T} \left [ {\rm e}^{i \gamma} 
+ |P/T| \, {\rm e}^{i \delta} \right ]~, 
\nonumber \\  
\sqrt 2 \, A^{00} & = &  
- |T| \, {\rm e}^{i \delta_T} \left [ |C/T| \, 
{\rm e}^{i \Delta} \, {\rm e}^{i \gamma} 
- |P/T| \, {\rm e}^{i \delta} \right ]. 
\nonumber 
\end{eqnarray} 
The charged-conjugate amplitudes  $\bar A^{ij}$ 
differ by the replacement  $\gamma \to - \gamma$. There are
5 dynamical parameters $|T|$, 
$r \equiv |P/T|$, $\delta$, $|C/T|$, $\Delta$, with 
$\delta_T = 0$ assumed for the overall phase. Thus, the
weak phase  $\gamma$ can be extracted together with other quantities
if the  complete set of 
experimental data on  $B \to \pi\pi$ decays is available, which  is not the case at present.   
%
%
%
%
  
Several isospin bounds have been obtained on the penguin-pollution angle
 $\theta$ (or $\alpha_{\rm eff}=\alpha+\theta)$~\cite{Grossman:1997jr,Charles:1998qx,Gronau:2001ff},
with the Gronau-London-Sinha-Sinha bound~\cite{Gronau:2001ff} being the strongest. These bounds are useful in
constraining the parameters of the $B \to \pi \pi$ system and have been used to reduce their
allowed ranges. The experimental branching ratios and the CP asymmetries
$A_{\rm CP}(\pi^+\pi^0)$, $A_{\rm CP}(\pi^+\pi^-)$ and $A_{\rm CP}(\pi^0\pi^0)$, as well as
the value of the coefficient $S_{\pi^+\pi^-}$ in time-dependent CP asymmetry have been fitted to determine the various
parameters. An updated analysis by Parkhomenko~\cite{Parkhomenko-ichep04} 
based on the paper~\cite{Ali:2004hb} yields the following values:
\begin{eqnarray}
\vert P/T \vert&=&0.51^{+0.10}_{-0.09};
~~\vert C/T \vert=1.11^{+0.09}_{-0.10};\nonumber\\
\delta &=& (-39.4 ^{+10.3}_{-9.8})^\circ;
~~\Delta = (-55.7 ^{+13.5}_{-12.5})^\circ ;\nonumber\\
\gamma &=& (65.3 ^{+4.7}_{-5.2})^\circ .\nonumber
\end{eqnarray}
 The range of $\gamma$ extracted from this analysis is in good agreement with the indirect
estimate of the same from the unitarity triangle. However,
the strong phases $\delta$ and $\Delta$ come out large; they are much larger
than the predictions of the QCD-F approach~\cite{Beneke:1999br} with
pQCD~\cite{Keum:2000ph,Hnli-ichep04} in better agreement with data, but 
neither of these approaches provides a good fit of the entire $B \to \pi \pi$ data.
Similar results and conclusions are obtained by Buras et al.~\cite{Buras:2004th}
and Pivk~\cite{Pivk-ichep04}.

Data on $B \to \pi \pi$  decays are in agreement with the phenomenological approach of
the so-called charming penguins~\cite{Pierini-ichep04}, and with the SCET-based analysis of
Bauer et al.~\cite{Bauer:2004tj}  which
also attributes a dominant role to the charming penguin amplitude. However, a proof of the factorization of the
charming penguin amplitude in the SCET approach remains to be provided.
 In addition, SCET makes a number of
predictions in the $B \to \pi \pi$ sector, such as
the branching ratio ${\cal B} (B^0 \to \pi^0\pi^0)$:
\begin{equation}
 {\cal B} (B^0 \to \pi^0\pi^0) 
\bigg |_{\gamma = 64^\circ} \!\!\! = 
( 1.3 \pm 0.6) \times 10^{-6}~, 
\end{equation}
which is in agreement with the current experimental  
 world average~\cite{Sakai-ichep04,Giorgi-ichep04}
\begin{equation}
 \bar {\cal B} (B^0 \to \pi^0\pi^0)  
= ( 1.51 \pm 0.28) \times 10^{-6}~.
\end{equation}
In contrast,  predictions of the QCD-F and pQCD approaches are rather similar:
 $ {\cal B} (B^0 \to \pi^0\pi^0) 
\sim  0.3 \times 10^{-6} $, in substantial disagreement with the data.

%
%
%


\subsection{$B \to K \pi$: SM vs.~Experiments}
The final topic covered in this talk is the $B \to K \pi$ decays. First, we note that the  direct CP-asymmetry in the $B \to K\pi$ decays has now been measured by
the BABAR and BELLE collaboration:
\begin{eqnarray}
A_{\rm CP}(\pi^+K^-) &=&\nonumber\\
&& \hspace*{-2.0cm} (-10.1 \pm 2.5 \pm 0.5)\%~~[{\rm BELLE}],\nonumber\\
&& \hspace*{-2.0cm} (-13.3 \pm 3.0 \pm 0.9)\%~~[{\rm BABAR}]~,
\end{eqnarray} 
to be compared with the predictions of the two
factorization-based approaches: $A_{\rm CP}(\pi^+K^-)= (-12.9 \sim -21.9)\%~[{\rm pQCD}]$~\cite{Keum:2000ph,Hnli-ichep04}
 and
$A_{\rm CP}(\pi^+K^-)= (-5.4 \sim +13.6)\%~~[{\rm QCD-F}]$~\cite{Beneke:1999br}, with the latter falling short of a
satisfactory description of data.

The charged and neutral $B \to \pi K$ decays have received a
lot of theoretical attention.
In particular, many ratios involving these decays have been
proposed to test the SM~\cite{Fleischer:1997um,Neubert:1998jq,Lipkin:1998ie,Buras:2000gc}
 and extract useful bounds on the angle $\gamma$, starting from the
Fleischer-Mannel bound~\cite{Fleischer:1997um} $\sin^2\gamma \leq R$, where the ratio $R$ is defined as follows:
\begin{equation}
R\equiv \frac {\tau_{B^+}}{\tau_{B_d^0}} 
        \frac{{\cal B}(B_d^0 \to \pi^-K^+) + {\cal B}(B_d^0 \to \pi^+K^-)}
         {{\cal B}(B^+ \to \pi^+K^0) + {\cal B}(B^- \to \pi^-\bar{K}^0)}~.    
 \end{equation}
The current experimental average $R=0.820\pm 0.056$ allows to put a bound: $\gamma < 75^\circ$
(at 95\% C.L.). This is in comfortable agreement with the determination of $\gamma$
from the $B \to \pi \pi$ decays, given earlier, and the indirect unitarity constraints. Thus, both
$R$ and $ A_{\rm CP}(\pi^+K^-)$ are in agreement with the SM. The same is the situation with the
Lipkin sum rule~\cite{Lipkin:1998ie}:
\begin{eqnarray}
 R_L&\equiv& 2 \frac{\Gamma (B^+ \to K^+\pi^0) + \Gamma(B^0 \to K^0\pi^0)}
{\Gamma (B^+ \to K^0\pi^+) + \Gamma(B^0 \to K^+\pi^-)}\nonumber\\
&=&1 +{\cal O}(\frac {P_{\rm EW} +T}{P})^2~;
\end{eqnarray}
implying significant electroweak penguin contribution
 {\it in case $R_L$ deviates significantly from 1}.
With the current experimental average $R_L=1.123 \pm 0.070$, this is obviously not the case.
This leaves then the two other ratios $R_c$ and $R_n$ involving the $B \to \pi K$ decays of 
$B^\pm$ and $B^0$ mesons:
\begin{eqnarray}
R_c &\equiv& 2 \big[ \frac{{\cal B}(B^\pm  \to \pi^0 K^\pm)}{{\cal B}(B^\pm  \to \pi^\pm K^0)}\big]~,
\nonumber\\
R_n &\equiv& \frac{1}{2} \big[ \frac{{\cal B}(B_d  \to \pi^\mp K^\pm)}{{\cal B}(B_d  \to \pi^0 K^0)}\big]~.
\end{eqnarray}
 Their experimental values $ R_c=1.004 \pm 0.084$ and $R_n=0.789 \pm 0.075 $ are to be
compared with the current SM-based estimates~\cite{Buras:2004th} $ R_c=1.14 \pm 0.05$ and $R_n=1.11 ^{+0.04}_{-0.05}$.
This implies $R_c({\rm SM}) - R_c ({\rm Exp})=0.14 \pm 0.10$ and
 $R_n({\rm SM}) - R_n ({\rm Exp})=-0.32 \pm 0.09 $. We conclude tentatively that SM is in agreement
 with the measurement of 
$R_c$, but falls short of data in $R_n$ by about 3.5$\sigma$. Possible deviations from the SM, if confirmed,
would imply new physics, advocated in this context, in particular, by Yoshikawa~\cite{Yoshikawa:2003hb},
Beneke and Neubert~\cite{Beneke:2003zv}  and  Buras et al.~\cite{Buras:2004th}

Finally, a bound on ${\cal B}(B^0 \to K^0 \overline{K^0})$
 based on $SU(3)$ and $B \to \pi \pi$ data, obtained recently by Fleischer and Recksiegel~\cite{Fleischer:2004vu},
yielding ${\cal B}(B^0 \to K^0 \overline{K^0})< 1.5 \times 10^{-6}$ is well satisfied by
the current measurements~\cite{Sakai-ichep04,Giorgi-ichep04} ${\cal B}(B^0 \to K^0 \overline{K^0})=(1.19\pm 0.38\pm 0.13) \times 10^{-6}$.

\vspace*{-0.3cm}
\section{Summary}
Dedicated experiments and progress in heavy quark
expansion techniques have enabled precise determination of the CKM matrix
elements entering in the unitarity triangle (\ref{eq:trianglerel}). In particular, 
 $\vert V_{cb} \vert$ is now determined
quite precisely: 
$\frac{\delta \vert V_{cb} \vert}{\vert V_{cb} \vert} \sim 2\%$
comparable to  $\frac{\delta \vert V_{us} \vert}{\vert V_{us} \vert}$.
  Current precision on  $\vert V_{ub} \vert$ from inclusive decays is about
 10\% and a factor 2 worse for the exclusive decays. There are 
several theoretical proposals to improve the knowledge of
 $\vert V_{ub} \vert$  requiring lot more data 
from the $B$ factories which will be available in the near future.

 The decay $B\to X_s \gamma$, which serves as the standard candle in the
FCNC $B$-decays,  is in agreement with the SM
with the current precision on the branching fraction at about 10\%. A major theoretical
effort is under way to complete the NNLO calculations in $B \to X_s \gamma$; at the same time 
digging deeper brings to the fore new hadronic  uncertainties which
will have to be controlled to reach the goal of 5\% theory precision in ${\cal B}(B\to X_s \gamma)$.
 Improved and new measurements in $B \to X_s \ell^+ \ell^-$ have been reported
including a first shot at the forward-backward asymmetry in $B \to K^* \ell^+ \ell^-$.
Data in the electroweak $b \to s \ell^+\ell^-$ sector is in agreement with the SM and this
 {\it rapport} will be tested  with greatly improved precision in the future.
 Current upper limit on ${\cal B}(B_s \to \mu^+ \mu^-)$ from CDF/D0 probes interesting SUSY parameter
space,  impacting on the large-$\tan \beta$ regime of SUSY.

Concerning non-leptonics, the
  largest current discrepancy between the data and the SM is in the decays involving QCD penguins.
 These include CP violation in the $b \to s \bar{s} s$ penguins, where data show a deviation of about
$3 ~\sigma$ from the SM~\cite{Sakai-ichep04,Giorgi-ichep04}. Also the ratio $R_n$ in the $B \to K \pi$ decays is
out of line with the SM-estimates by slightly more than 3 $\sigma$. These deviations are not yet
 significant enough
to announce physics beyond the SM; neither can they be wished away.
  Experimental evidence is now mounting  that
not only are the weak phases $\alpha,~~\beta,~~\gamma$ large, as anticipated in the SM, 
but so also are the  strong (QCD) phases, unlikely to be generated by perturbative QCD alone.
In addition, 
color-suppressed decays are not parametrically suppressed, as opposed to their estimates in
the QCD-F and pQCD approaches. 
 SCET-- the emerging QCD technology -- holds the promise to provide a better theoretical
description of non-leptonics than existing methods.
We look forward to theoretical developments as well as to
new and exciting data from the ongoing and planned experiments
at the $B$ factories and hadron colliders.

{\bf Acknowledgments}

I would like to thank Christian Bauer, Geoffrey Berryhill, Andrzej Buras, Thorsten Feldmann, Robert Fleischer, Paolo Gambino,
 Christoph Greub, Yong-Yeon Keum,
 Alexander Khodjamirian, Hsiang-nan Li, Harry Lipkin,
 Cai-Dian Lu, Vera L\"uth, Enrico Lunghi, Thomas Mannel, Mikihiko Nakao, Matthias Neubert,
Jan Olsson, Dmitri Ozerov, Alexander Parkhomenko, and Kolya Uraltsev for many discussions and valuable input. I also thank my
wife Lyuba Vasilevskaya for her help with the preparation of this talk and her support.

\end{document}